\documentclass[twocolumn,aps,prl,superscriptaddress,showpacs]{revtex4-1}
\usepackage{graphicx}
\usepackage{dcolumn}
\usepackage{bm}
\usepackage{color}
\usepackage{amsmath}
\usepackage[english]{babel}

\begin{document}

\title{Phase and Amplitude single-shot measurement by using heterodyne
  time-lens and ultrafast digital time-holography }

\author{Alexey Tikan, Serge Bielawski, Christophe Szwaj, St\'ephane Randoux and Pierre Suret}

\affiliation{Laboratoire de Physique des Lasers, Atomes et Molecules,
  UMR-CNRS 8523,  Universit\'e de Lille, France \\
Centre d'Etudes et de Recherches Lasers et Applications (CERLA), 59655 Villeneuve d'Ascq, France}


\maketitle



{\bf Temporal imaging systems are outstanding tools for single-shot observation of optical signals that have irregular and ultrafast dynamics. They allow long time windows to be recorded with femtosecond resolution, and do not rely on complex algorithms. However, simultaneous recording of amplitude and phase remains an open challenge for these systems. Here we present a new heterodyne time-lens arrangement that efficiently records both the amplitude and phase of complex signals, while keeping the performances of classical time-lens systems ($\sim 200$~fs) and  field of view (tens of ps). Phase and time are encoded onto the two spatial dimensions of a camera. We demonstrate direct application of our heterodyne time lens to turbulent-like optical fields and optical rogue waves generated from nonlinear propagation of partially coherent waves inside optical fibres. We also show how this phase-sensitive time-lens system enables digital temporal holography to be performed with even higher temporal resolution (80 fs).}\\

Simultaneous measurement of the amplitude and phase of ultrafast
complex optical signals is a a key question in  modern  optics and
photonics~\cite{reid2016roadmap,Trebino,Broaddus:10,bowlan2006crossed,alonso2010spatiotemporal,
  Rhodes:14}. This kind of detection is needed for the
characterization of various fundamental phenomena such as  e.g.
supercontinuum   \cite{Wetzel:12,Wong:14}, optical rogue waves (RWs)
\cite{Solli:07,Suret:16,Narhi:16}, or  soliton dynamics in mode-locked
lasers \cite{Herink:17, Ryczkowski:17}. The task remains a particulary
challenging open problem when femtosecond resolution and long time
windows  are simultaneously required. These requirements are found for exemple in the context of nonlinear statistical optics and of the characterization  of random light ~\cite{Picozzi:07} or in the study of spatio-temporal dynamics of lasers~\cite{Turitsyna:13}.

In the quest for long-window and ultrafast recording tools, temporal imaging devices, such as time-lenses, are considered as promising candidates. Time-lenses enable femtosecond time evolutions to be manipulated so that they can be magnified in time~\cite{Kolner:89,Foster:08,Bennett:99} or spectrally encoded~\cite{Kauffman:94,Foster:08,Suret:16} with high
fidelity. These signals evolution replica can thus be recorded using a simple
GHz oscilloscope (for time-magnification systems) or a single-shot optical
spectrum analyzer~\cite{Suret:16}). No special algorithms are necessary
for retrieving the ultrafast power evolutions, long windows can be recorded
(up to hundreds of picoseconds), and the method is suitable for recording
continuous-wave (i.e., non-pulsed) complex signals. Recently, temporal imaging
systems have  thus begun to play a central role in fundamental studies dealing with
nonlinear propagation of light in fibers leading for example to the emergence of rogue waves and integrable turbulence~\cite{Suret:16,Narhi:16,fridman2017measuring},
where recording long temporal traces with femtosecond resolution is
mandatory. Commercial devices are also available in the market (by PicoLuz
LLC).

However, a range of applications is still hampered by the need to also record
the phase evolution of long and complex ultrafast optical signals. Extension
of temporal imaging has been performed in this direction, by performing
heterodyning in temporal magnification systems
~\cite{Dorrer:06,dorrer2008electric}, i.e., for which the readout is
performed using a single-pixel photodetector. However these systems have not
demonstrated sub-picosecond capability.

 Here we show that temporal imaging systems can be transformed in a full amplitude and phase digitizing device, without major trade-off on femtosecond resolution and recording window. The principle is to use a spectral-encoding time-lens, which encodes the time
evolution onto the horizontal axis of a camera, and obtain the phase
information by performing heterodyning on the other (vertical) direction. This spatial encoding arrangement reminds strategies used in two-dimensional interferometry,
SEA-TADPOLE~\cite{bowlan2006crossed}, SEA-SPIDER~\cite{Kosik:05,Wyatt:06} and STARFISH~\cite{alonso2010spatiotemporal}, except that we image directly the time-evolution (instead of a spectrum) on the camera.

After presenting the experimental arrangement, we will  consider three examples of applications of our heterodyne time microscope (HTM). (i) First, as a test of the phase
  measurement capability, we will present direct recordings of the continuous
  wave produced by a narrowband laser. (ii) Then we will present recordings of the
  light produced by a partially coherent amplified spontaneous emission (ASE)
  source. We will also present recordings of optical signals emerging from
  the nonlinear propagation of such a random light in optical
  fibres. In particular, by providing the phase and amplitude signatures, we give the awaited experimental proof that  the Peregrine soliton (PS) spontaneously
    emerge locally in nonlinear random wave trains~\cite{Bertola:13,Suret:16}.  (iii) Eventually, we will show how the method can be straighforwardly   extended to perform digital temporal holography. Remarkably, by avoiding the aberrations induced by high order dispersion of usual time lens devices \cite{Bennett:00,Bennett:01,Salem:13}, the effective resolution of the time holography is found to
be  around  $80$~fs.

\section*{Results}

\subsection{Experimental strategy}

\begin{figure*}[th!]
\includegraphics[width=18cm]{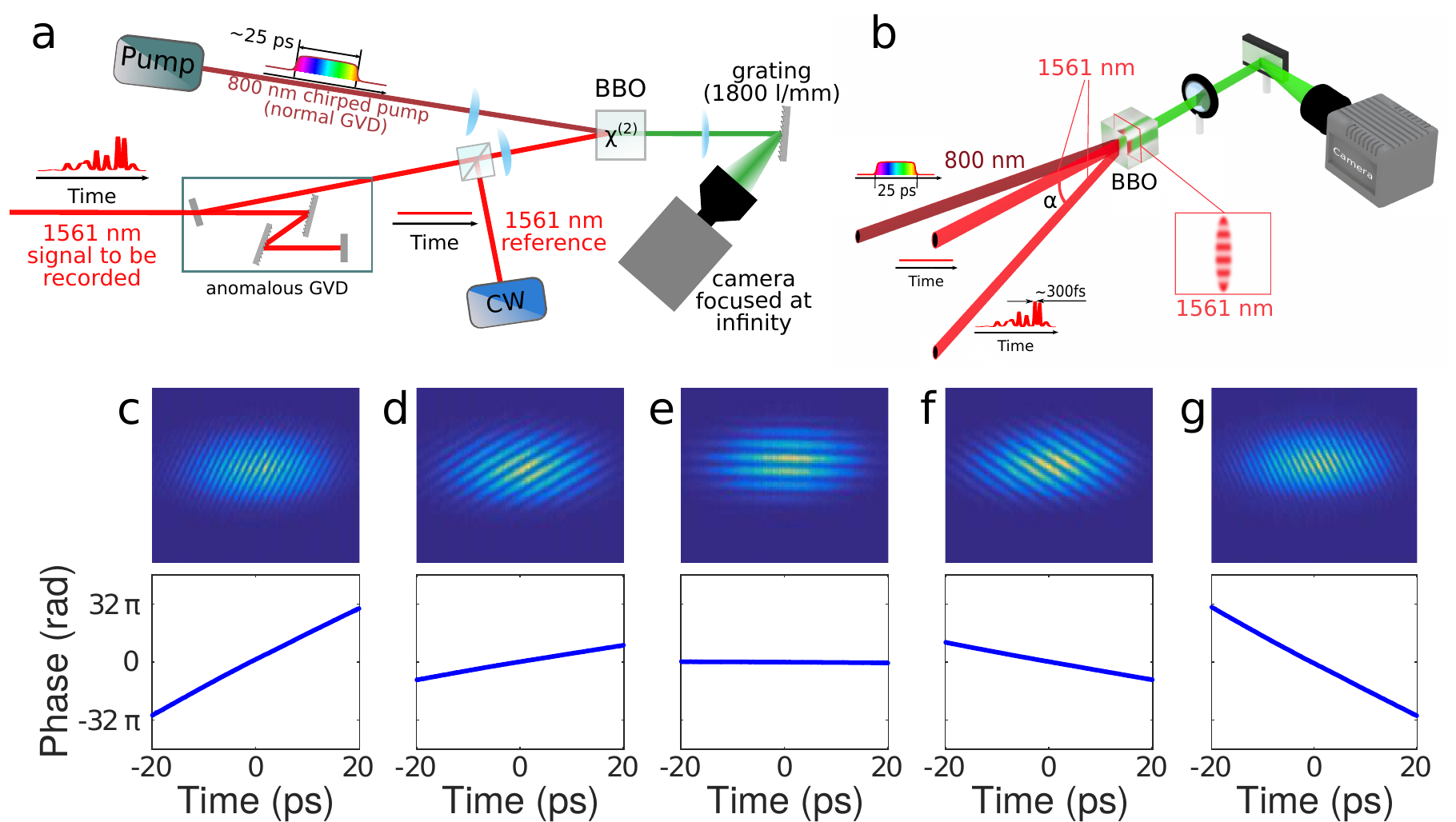}
\caption{{\bf Heterodyne time-lens arrangement for amplitude
      and phase measurement of arbitrary signals.} (a) Top view of the whole
    setup. Without the reference CW signal, the system is a time-microscope,
    which produces a ``spatial replica'' of the signal power evolution (at
    $\lambda_S \sim 1560$~nm) onto the horizontal axis of the camera (the replica has 529~nm wavelength). The noncolinear Sum Frequency Generation (SFG) between the optical field at $\lambda_S$ and a pump chirped pulse at $\lambda_P \sim 800$~nm  in the BBO crystal plays the role of a ``time-lens'' for the $\approx$1560~nm signals (see text).
    Interference with the reference CW local
    oscillator (at $\lambda_R$ near $\lambda_S$) adds interference fringes
    along the vertical direction of the camera, which are used to retrieve the
    phase information.  In the crystal, all the beams are elliptic with horizontal and vertical diameters $ \sim 50\mu$m and $\sim 500 \mu$m respectively. Note that all optical paths are horizontal, except the
    CW local oscillator signal [which enters the BBO crystal with a small angle $\alpha \sim 10$~mrad with respect to the horizontal plane, as shown in (b)].  The sub-system composed by the grating and the sCMOS camera forms a    single-shot Optical Spectrum Analyzer (OSA) for the 529~nm SFG output.  (c-g) Raw images and retrieved phase evolutions obtained when the signal
    comes from a monochromatic source with various wavelengths: $\lambda_S =1555.37$~nm (c), $\lambda_S =1559.37$~nm (d), $\lambda_S =1561.37$~nm (e),  $\lambda_S =1563.37$~nm (f) and $\lambda_S =1567.37$~nm (g). In all case the wavelentgh of the reference is $\lambda_R =1561.37$~nm. For clarity, transport optics and filtering elements at BBO output have been omitted, see Sec Methods.\\ }
\label{fig:1}
\end{figure*}

Our experimental setup for the measurement of phase and amplitude is
 displayed Figure \ref{fig:1}.(a) and a detailed 3D enlarged view of
 the heterodyne time lens is plotted in  Fig. \ref{fig:1}.(b).

The time lens ({\it i.e.} the temporal quadratic phase
\cite{Kolner:89}) is provided by noncolinear Sum Frequency Generation
(SFG) inside a $\chi^{(2)}$ BBO crystal between the signal field at
$\lambda_S \sim 1560$ nm and a chirped pump pulse at $\lambda_P \sim 800$~nm. Measurement of the optical phase is achieved from an heterodyne setup based on the beating of the signal beam with a reference single-frequency laser emitting at a fixed wavelength
$\lambda_R = 1561$~nm. The reference beam reaches the BBO crystal by
making a small vertical angle $\alpha \sim 10$~mrad with respect to
the incidence plane that contains the pump beam and the signal beam. This produces an interference
pattern at $\lambda \sim 1561$~nm that is copied  at
$\lambda_{vis}\sim 529$~nm by the process of SFG.

The spectrum of the light generated at $\lambda_{vis}$  is imaged onto the horizontal axis of the camera by using a simple diffraction grating. In the horizontal direction, the device is similar in its principle to the time microscope reported in \cite{Suret:16} where time is encoded into space : the horizontal position on the camera corresponds to the temporal evolution.  Note that the horizontal (time) axis was calibrated  from annex experiments involving double pulses (see Sec. Methods).

 In the vertical direction, the position of the fringes of interference ({\it i.e.} for example the position of the maxima of the interference pattern) is directly proportionnal to the relative phase between the signal under investigation and the monochromatic reference field (see Sec.~Methods). As a consequence, it is straightforward to extract the phase information from the pattern recorded in single shot by the sCMOS camera : the horizontal change in the vertical position of the fringes of interferences directly provides the ultrafast evolution of the phase of the signal. Our heterodyne time microscope (HTM) provides single-shot snapshots of the phase and the amplitude of the signal over time windows having a width around  $40$~ps. Those single-shot recordings are performed synchronously with the pump laser, at 1~KHz repetition rate (see Sec. Methods).\\

\subsection{Tests using a continuous-wave source}

In order to test our HTM and to illustrate its operating principle, we
first examine the phase evolution $\phi_S(t)$ of a monochromatic signal having a
wavelength $\lambda_S$ that can be tuned over some wavelength range
(see Sec.~Methods). The beating pattern observed on the camera is
plotted in Figs.~\ref{fig:1}.(c-g) for five different values of
$\lambda_S$. Note that the wavelength $\lambda_R$ of the reference
laser is kept fixed. One immediatly sees that the fringes are straight
lines whose slope depends on the pulsation difference $\delta \omega$
between the  pulsation of the signal light and the one of the reference source. The phase $\phi_R$ of the reference laser can be considered
as constant over each $\sim 40$~ps-long temporal window of measurement
(see Sec. Methods). The relative phase $\phi(t)=\phi_S(t)-\phi_{R}$ is
simply determined by using a procedure similar to the one described in \cite{Kreis:86} for each value of the time $t$, {\it i.e.} for each vertical line of the pattern (see Sec.~Methods).

As expected we find that the phase obeys a linear time evolution $\phi(t)=\delta \omega  \, t$  (see Fig.~\ref{fig:1}.(e-g)). In the experiments reported in Figs.~\ref{fig:1}.(c-g), instantaneous frequencies as large as $\delta \omega = \pm 0.8$~THz have been observed. 


\subsection{Application to the analysis of a partially coherent light source}

We now demonstrate how our device can be used to investigate phase and
amplitude fluctuations of random light. Contrary to the previous
situation where the signal was monochromatic, the time microscope must
now be carefully set in a situation where the distance between the
object under investigation and the objective lens is adjusted in such
a way that a well-defined  image is observed on the camera. This is
achieved by carefully tuning the dispersion experienced by the random
signal light by the use of a Treacy grating compressor (see
Fig.~\ref{fig:1}.(a), Sec. Methods and ~\cite{Kolner:89,Bennett:99, Foster:08, Suret:16}).

\begin{figure*}
\includegraphics[width=16cm]{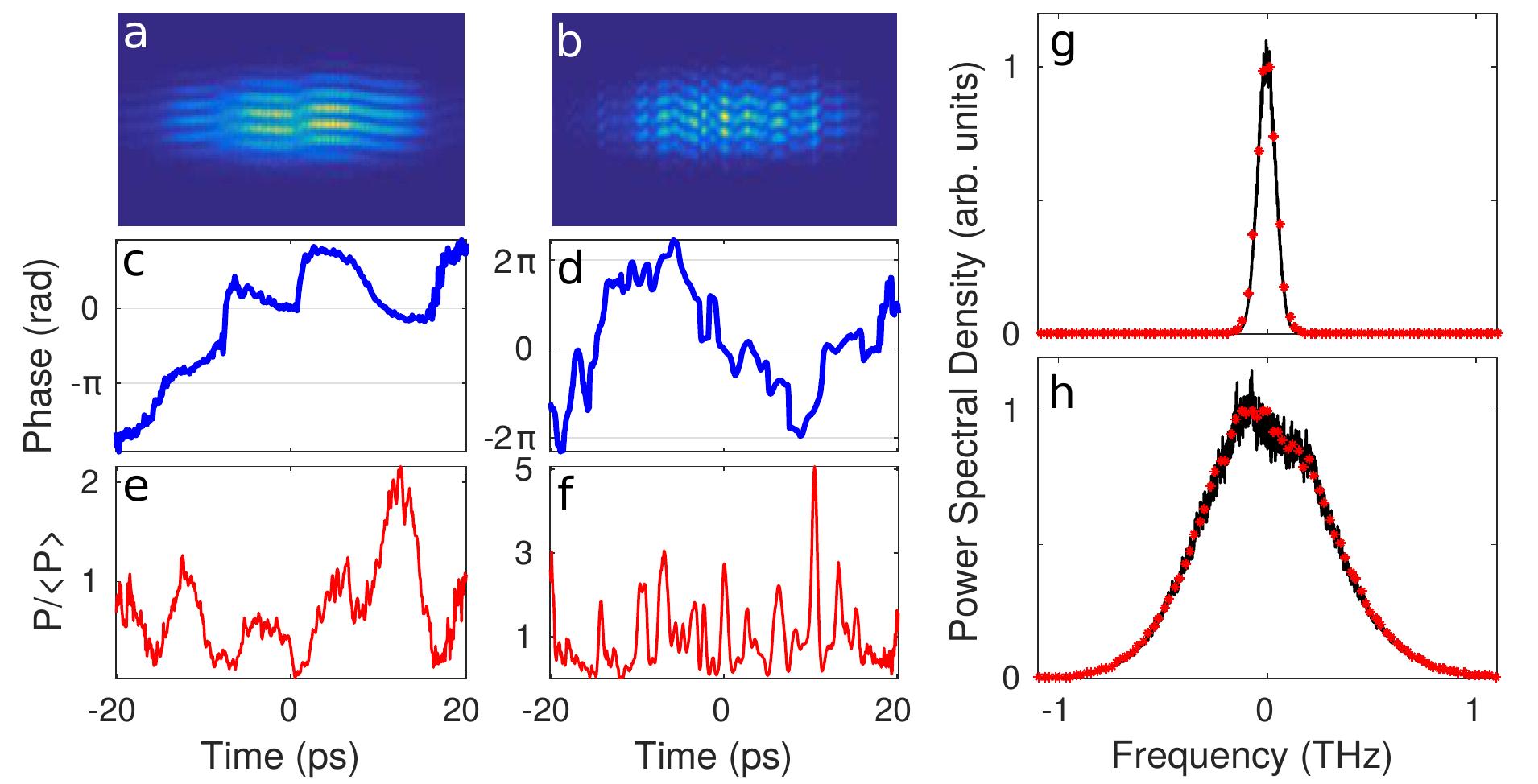}
\caption{{\bf Phase, amplitude and spectrum of partially coherent waves (ASE).} (a) and (b), Typical raw images recorded by the sCmos camera of the HTM. (c) and (d)  phase retrieved from the pattern. (e) and (f) optical power normalized to the average power. Note that on fig a and b, the signal is not divided by the averaged envelop; on the contrary on the Figs. e and f the signal is divided by the average power ($P/\langle P\rangle$). The spectral width is $\Delta \nu=0.1$~THz for (a), (c), (e) and $\Delta \nu=1.$~THz for (b), (d), (f).  (g) and  (h) :  {\bf optical spectra} corresponding to  $\Delta \nu=0.1$~THz  and  $\Delta \nu=1.$~THz respectively. Red crosses represent the spectrum computed from the averaged Fourier transform of the envelop of the electric field recorded with the HTM. Comparison with the spectrum recorded with an OSA (black line).}
\label{fig:2}
\end{figure*}

We first investigate the partially coherent waves  emitted by an ASE light source at a wavelength $\lambda\sim1561$ nm (see Sec. Methods).  Using a programmable optical filter, the optical spectrum of the partially coherent light is designed to assume a Gaussian shape with a full width at half maximum $\Delta \nu$ that can be adjusted to some selected
values.

  Typical 2D patterns recorded in single shots by
using the camera are displayed in Fig.~\ref{fig:2}.(a,b) while the
retrieved phase $\phi(t)$ and the optical power  fluctuations  $P(t)$ are displayed in
Figs.~\ref{fig:2}.(c,d) and \ref{fig:2}.(e,f) for $\Delta \nu=0.1$ THz
and $\Delta \nu=1$ THz respectively. Note that the retrieval algorithm
is very simple and straightforward. For each value of $t$, the phase $\phi(t)$ is given by the position of the maxima and the optical power fluctuations  $P(t)$ is simply computed from sum of along vertical lines of the 2D patterns and from the knowledge of the reference power previously recorded (see Sec. Methods).

As it can be expected, both the phase and of the power of the ASE light randomly fluctuate over time scales of the order of $1/\Delta \nu \sim 10$ps [Fig.~\ref{fig:2}.(c,e)] and $1/\Delta \nu \sim 1$ps [Fig.~\ref{fig:2}.(d,f)]. From additionnal experiments with pulsed laser, the temporal resolution of our heterodyne time microscope is found to be around $200$~fs.\\

Furthermore, a stringent test of the HTM can also be performed by comparing the optical spectrum deduced from the data, with the optical spectrum recorded independently using an optical spectrum analyzer. For each frame (recorded every ms), we compute the Fourier Transform $\widetilde{A(\omega)}$ of the complex envelop $A(t)=\sqrt{P(t)} \, e^{i \phi(t)}$ of the electric field.  $P(t)$ and $\phi(t)$ represent the power and the phase of the ASE light measured by our HTM respectively. Secondly, the mean Fourier power spectrum
$\langle\widetilde{|A(\omega)|^2}\rangle$  of the partially coherent light is computed from an
average ensemble that is made over 5$\times 10^4$ frames. The spectral power  densities
computed with this procedure for $\Delta \nu=0.1$THz and $\Delta
\nu=0.1$THz are plotted in red crosses on the Figs. \ref{fig:2}.(g)
and \ref{fig:2}.(h) respectively. In both cases, the spectral power
density of the ASE light separately recorded with an optical 
spectrum analyser (OSA) is plotted in black lines. The agreement between the
spectra computed from the data recorded by the HTM and the
 spectra recorded with the OSA is
remarkably good. These experiments with ASE light demonstrate that our HTM
  is indeed able to perform the accurate single-shot simultaneous measurement of phase and power of partially coherent waves fluctuating over subpicosecond time scales.\\ 

\subsection{Application to nonlinear random waves : optical rogue
  waves and Peregrine soliton in integrable turbulence}

Now, we use the specific abilities of our HTM to investigate complex
spatio temporal structures that emerge from nonlinear propagation of
partially coherent waves inside optical fibers. For the sake of
simplicity, we restrict our study to the framework of the so-called
{\it integrable turbulence} \cite{Agafontsev:15, Walczak:15,
  SotoCrespo:16,Suret:16, Randoux:17}. This field of research, first
introduced by Zakharov, deals with statistical properties of wave
systems that are described by integrable equations such as the
one-dimensional nonlinear Schr\"odinger equation
(1D-NLSE)~\cite{Zakharov:09}.

\begin{figure*}
\includegraphics[width=16cm]{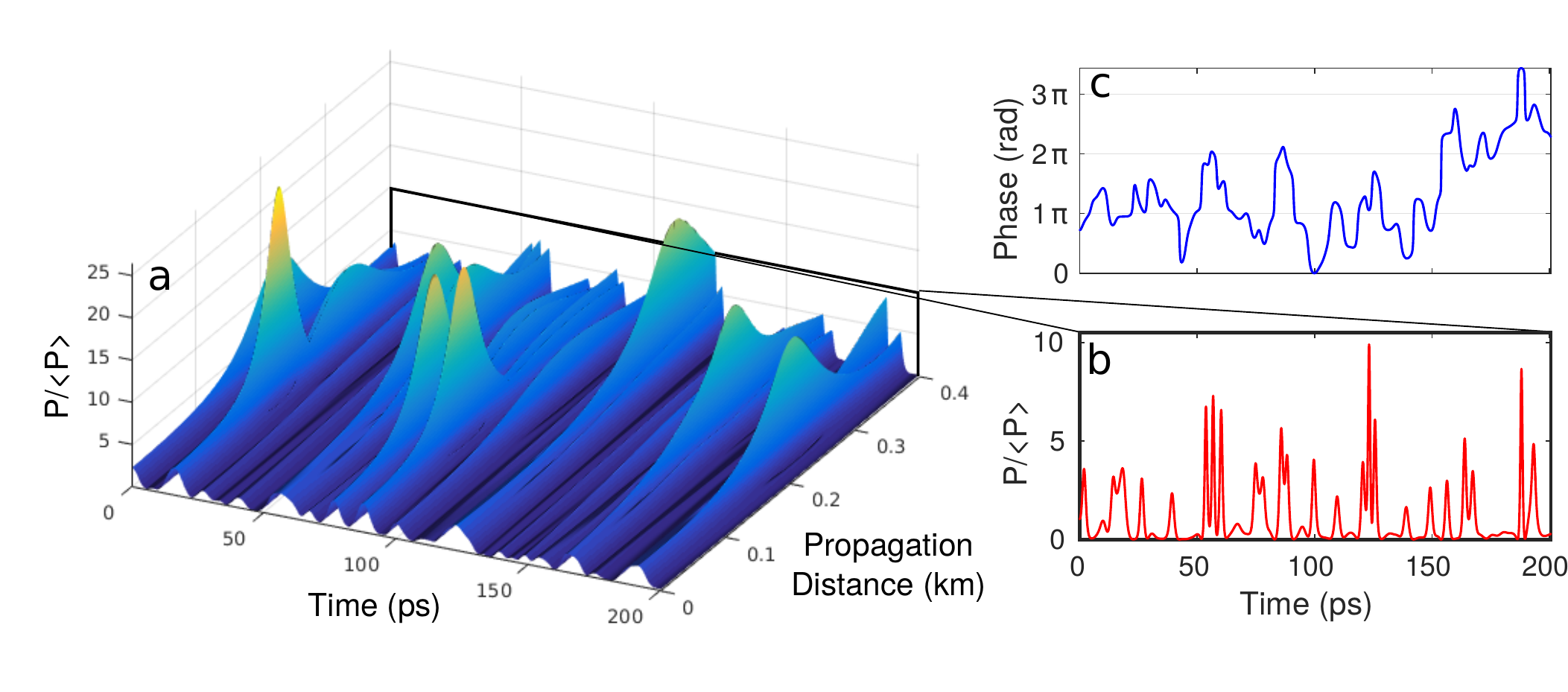}
\caption {{\bf  Numerical simulations of 1D-NLSE} (a) Spatiotemporal evolution of partially coherent waves in a 400~m-long polarization maintaing fibre (PMF). The initial condition have the same spectral and statistical properties as those of the ASE light used in the experiments [See Fig.~\ref{fig:2}.a]. The parameters correspond to those of the experiments [see Fig.~\ref{fig:4}].  (b) optical power and (c) phase of  partially coherent waves after  the nonlinear propagation.}
\label{fig:3}
\end{figure*}

First, for the sake of comparison with the experiments, we perform
numerical simulations of 1D-NLSE in the context of integrable
turbulence (see Fig.~\ref{fig:3}). The initial conditions used
in numerical simulations have the same spectral and statistical
properties as the ASE light previously considered
[Fig.~\ref{fig:2}.a)]. The parameters used in numerical simulations
correspond to the experiments that are described below (see Sec.
Methods).  It is known that the emergence of rogue waves that are shown in Fig.~\ref{fig:3}.a) are responsible  for deviations from  the initial gaussian statistics~\cite{Walczak:15,Suret:16,SotoCrespo:16}. Fig.~(\ref{fig:3}.b) show typical power and phase dynamics
 computed after the nonlinear propagation. Up to now, it was not possible to observe this ultrafast dynamics of the phase of optical rogue waves.

\begin{figure*}
\includegraphics[width=16cm]{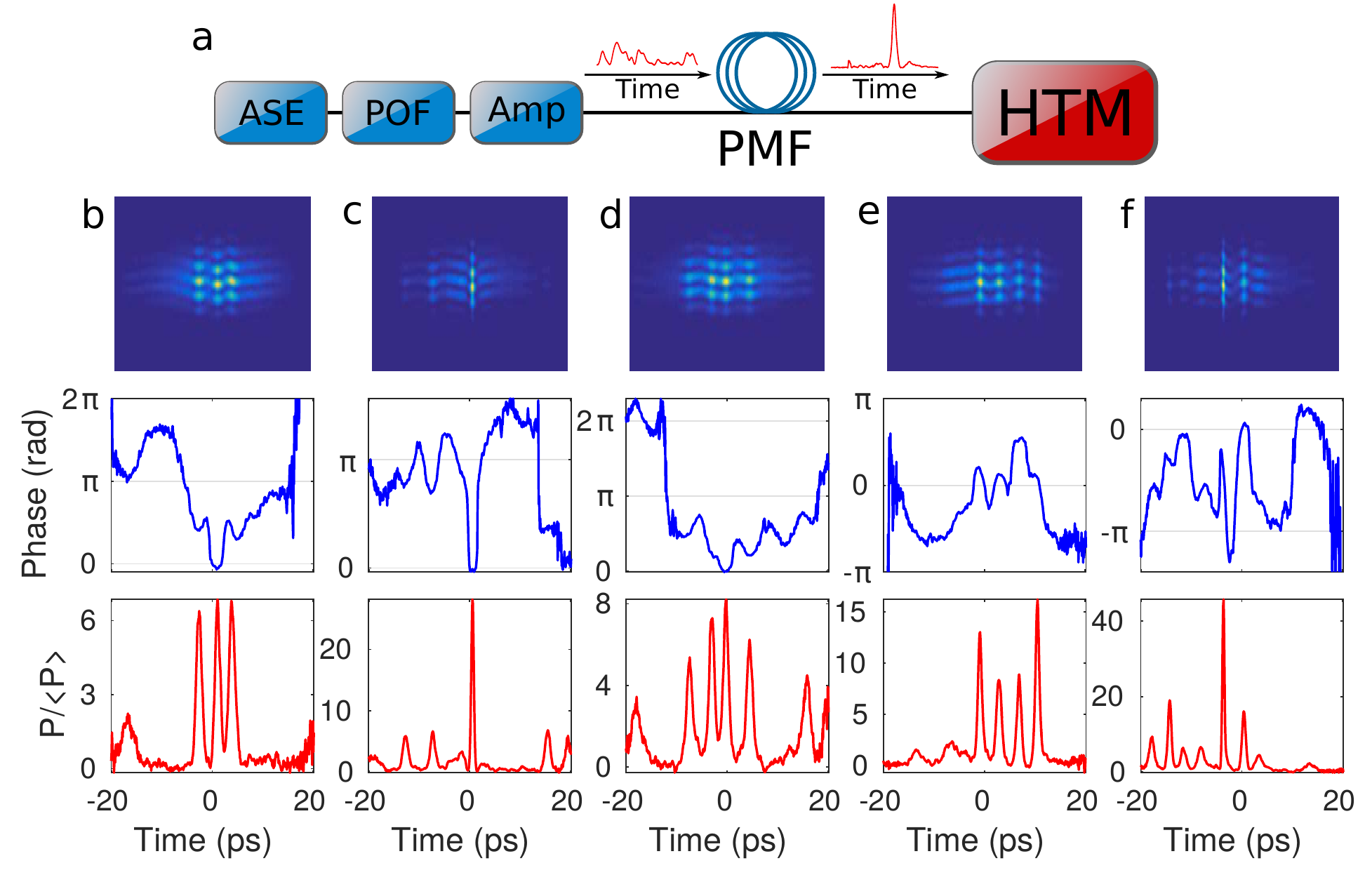}
\caption {{\bf Experiments on nonlinear random waves} (a) Experimental
  setup. ASE: Erbium fiber broadband Amplified Spontaneous Emission
  source. POF: Programmable Optical Filter to tailor the optical spectrum of the initial condition. AMP: Erbium-doped   fiber Amplifier. PMF: 400~m- ong Polarization Maintaining Fibre. HTM: Heterodyne Time Microscope of  Fig.~\ref{fig:1}. (b-f) Typical raw images (top) retreived phase  (middle) and optical power (bottom) of partially coherent waves
  after nonlinear propagation in a 400m-long PMF. Note that (c)
  display a structure similar to the Peregrine Soliton. The initial spectral width of partially  coherent waves is $\Delta \nu=0.1$THz.  The Average Optical power launched into the PMF is $\langle P \rangle=2.6$W.}
\label{fig:4}
\end{figure*}

In order to investigate experimentally this complex phase and
amplitude dynamics, we have launched partially coherent waves 
emitted by the ASE light source (see above and Sec. Methods) into 
$100$~m- or $400$~m-long Polarization Maintaining Fibres (PMF) at a wavelength 
falling in the anomalous (focusing) dispersion regime. The average power of the ASE light in the PMF is $\langle P \rangle \simeq 2.6$W.  Fig.~\ref{fig:4}
represents the typical time evolutions of phase and amplitude that are
observed at the output of the  PMF by using our HTM. As mentioned above and already reported in \cite{Walczak:15,Suret:16}, the optical power fluctuations exhibits extreme events (RWs) having large amplitude with short time scale ($\sim 500$~fs). Remarkably, the dynamics of the phase and of the amplitude exhibit the same qualitative features in numerical simulations [Fig.~\ref{fig:3}.b] and in experiments [Fig.~\ref{fig:4}].\\

Single-shot recording of the phase is crucial for testing theoretical predictions of nonlinear statistical optics. This requirement is particularly stringent for unambiguous conclusions about the nature of coherent structures emerging from nonlinear random waves. As an example of the strength of our new technique, we study here the  scenario implying the local formation of the Peregrine soliton
(PS) that has been proposed to explain the emergence of heavy-tailed
statistics~\cite{Bertola:13,Suret:16, Tikan:17}.

\begin{figure*}
\includegraphics[width=15cm]{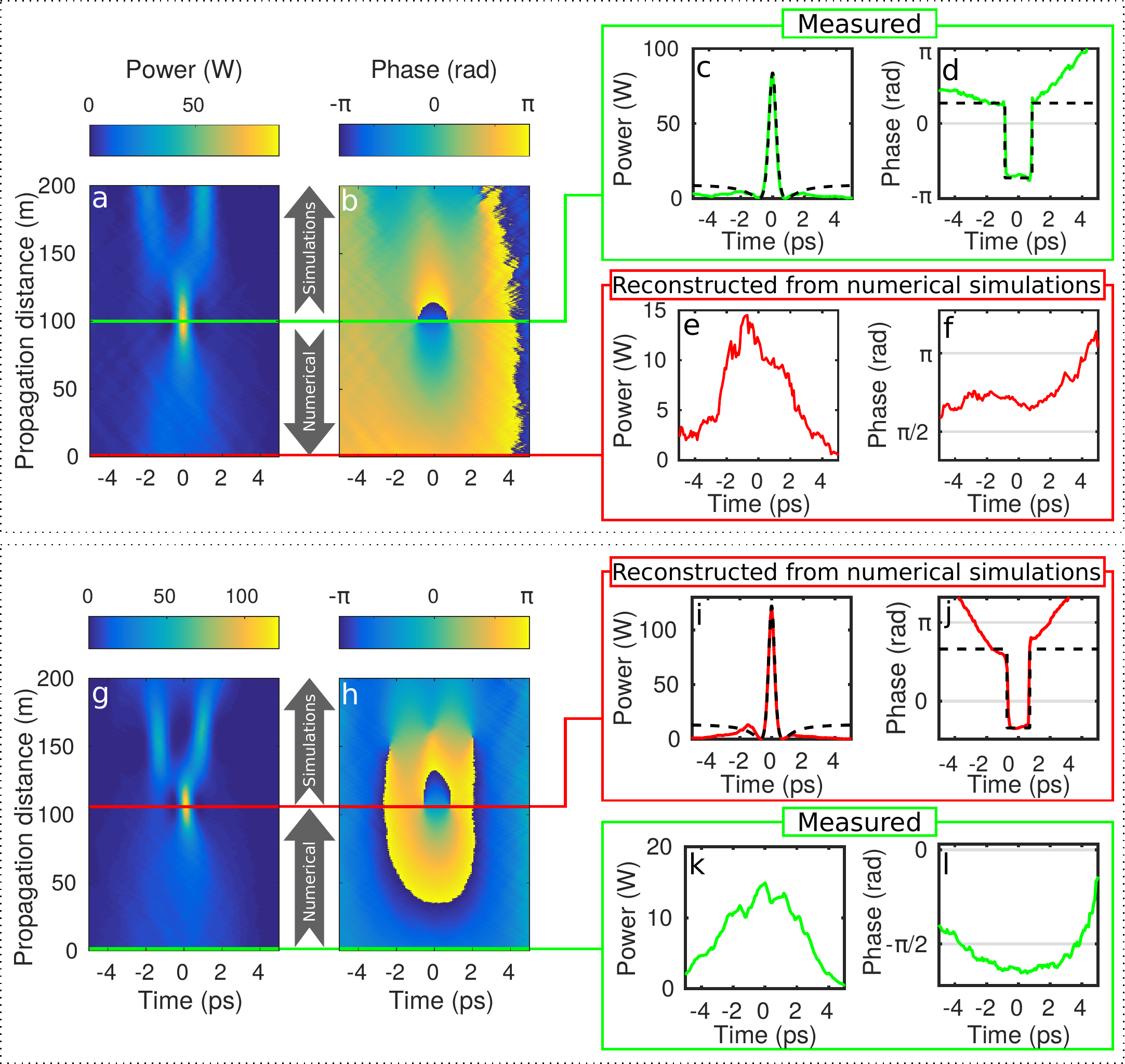}
\caption {{\bf Nonlinear Holography : Numerical simulations of 1D-NLSE
    from experimental data}. Top (a-f): Numerical simulation using as initial condition (green line on (a) and (b)) the experimental data recorded at the output of the 100~m PMF. Simulations are made forward (from 100~m to 200~m) and backward (from 100~m to 0~m). Spatiotemporal evolution of the power (a) and phase (b). The green inset represent plots versus time of the power (c) and  phase (d) of the experimental data used as initial condition. The red inset represent plots versus time of the power (e) and phase (f) obtain after numerical simulation of the backward propagation at 0~m (red line on (a) and (b)). Dashed line: Analytical solution of the Peregrine Soliton. Bottom (g-l): same as (a-f) but the numerical simulation is made forward (from 0~m to 200~m). The green inset: initial experimental condition. Red inset: numerical result at 100~m.}
\label{fig:5}
\end{figure*}

The temporal evolution of the power and of the phase plotted in
Fig.~\ref{fig:4}.(b) shows typical structures that looks very much
like the PS. Another similar structure is plotted in green line in
Figs.~\ref{fig:5}.(c,d) together with plots made from the analytical formula characterizing the PS (black dashed line, see \cite{Kibler:10}). The local intensity and phase profiles experimentally measured coincide well with those typifying the  PS. In particular, one observes the characteristic $\pi$ phase jump at times where the optical power falls to zero. This observation confirms the conclusion of previous studies that the PS represents a coherent  structure of special importance in the context of integrable turbulence \cite{Suret:16,Narhi:16}. However, by providing temporal signatures of both the phase and power profiles, our experiments definitively demonstrate that the PS can locally emerge from nonlinear random waves. This result deserves to be connected with some theoretical and experimental results which have shown that the PS is the coherent structure that is produced from the regularization of gradient catastrophes occurring in deterministic (not random) strongly nonlinear regime~\cite{Bertola:13, Tikan:17}.\\

The knowledge of the phase and of the amplitude of the experimental
field  now opens the way to a full comparison between experiments and
theory. In particular, as previously proposed in the context of experiments involving short pulses propagating in optical fibres \cite{Tsang:03} and of 2D experiments in
photorefractive crystals \cite{Barsi:09}, we can perform nonlinear digital
holography. The idea is to use phase and amplitude
profiles measured in the experiments as initial conditions in
numerical simulations of 1D-NLSE. Numerical simulations can be
performed either in the backward (reverse) direction or in the forward
direction (for longer propagation than in the experiments).
This technique provides the spatio-temporal dynamics inside the
optical fibre.

We first use experimental data recorded at the output of the $100$~m-long PMF (green lines in Figs.~\ref{fig:5}.(a,b,c,d)) as initial condition in our numerical simulations. We have integrated numerically the 1D-NLSE from $z=100$~m to $z=0$~~m and  from $z=100$~m to $z=200$m. The spatiotemporal evolution of the phase and amplitude obtained from numerical simulation are shown in Figs.~\ref{fig:5}.(a,b). The power and phase profiles computed from the 1D-NLSE at $z=0$~m [see red line in Fig.~\ref{fig:5}.(e,f)] reveals a good qualitative agreement with time fluctuations of the partially coherent waves  launched inside the PMF and recorded with the HTM [see green line in Fig.~\ref{fig:5}.(k,i)]. In particular, the field computed at $z=0$ m exhibits fluctuations having the expected time scale ($\sim 10$~ps). The numerical integration of the 1D-NLSE between $z=100$~m and $z=200$m shows that the PS-like structure observed in our experiment would split into two sub-pulses if nonlinear propagation was prolongated over another $100$-m long PMF.

Now, the phase and amplitude profiles that have been measured for the ASE light [green line in Fig.~\ref{fig:5}.(k,i)] are used as initial conditions for the numerical integration of the 1D-NLSE between $z=0$~m to $z=200$m [see Figs.~\ref{fig:5}.(g,h)].  Numerical simulations reveals the growth of a PS-like structure that splits into two separate pulses after $z \sim 100$ m. Note that this spatio-temporal features are very similar to those typifying nonlinear propagation of the so-called N-solitons \cite{Yang,Agrawal} but here, it is found  in the context of the propagation of partially coherent waves.\\

\begin{figure*}
\includegraphics[width=18cm]{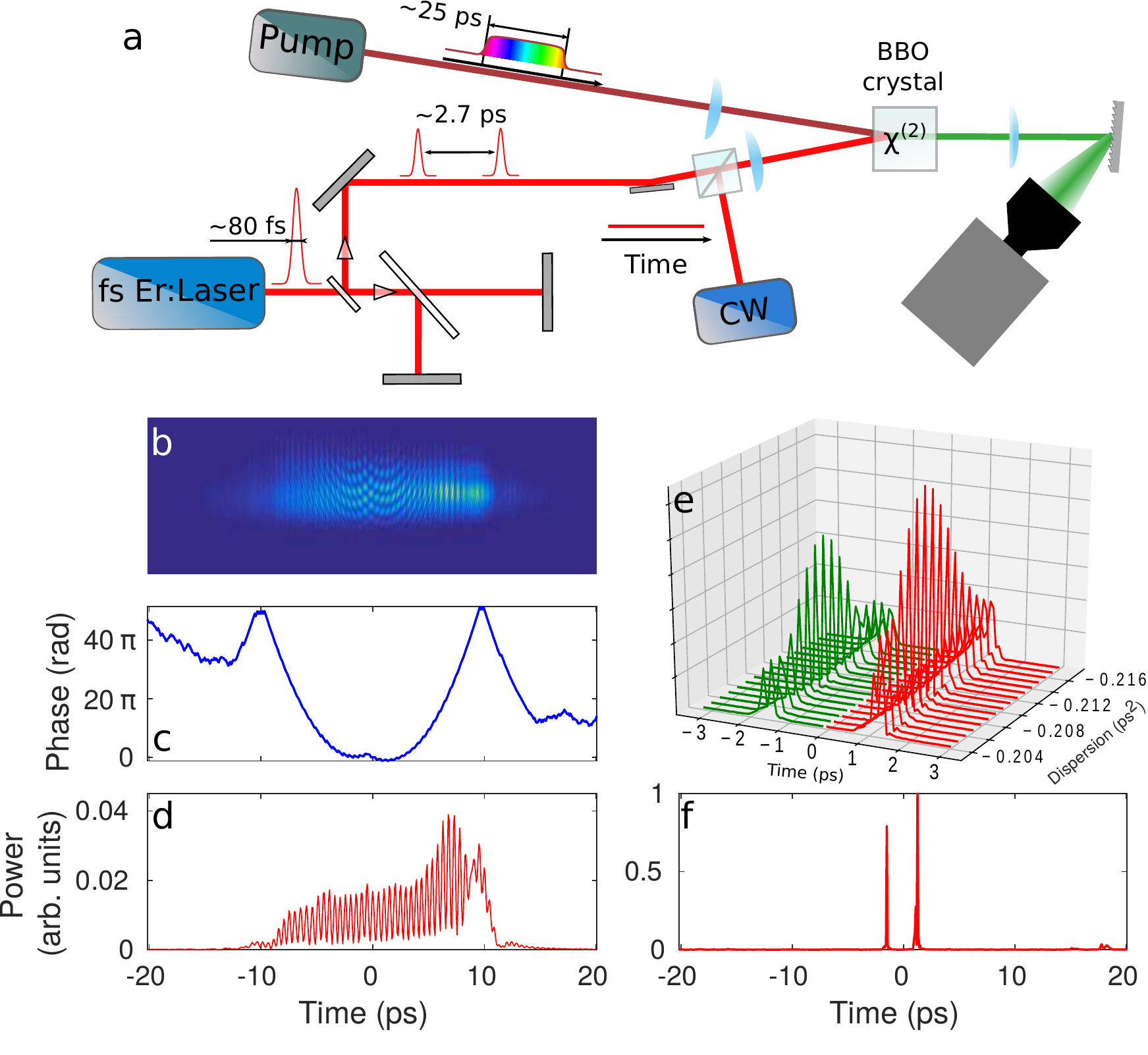}
\caption{{\bf Digital time-holography setup (SEAHORSE), and test data}. The
    setup displayed in (a) is similar to the one of Figure~\ref{fig:1}, except
    that no dispersion is induced on the signal path. In this test, we use a
    signal that is composed of two successive pulses with known delay. (b) Raw
    single-shot camera image, (c) and (d) corresponding evolutions of the    phase and power (see text for their precise meanings). (f) input signal
    retrieved from (c,d) using a digital holography algorithm (see text). (e)
    signal shapes versus digital focus adjustment.}
\label{fig:6}
\end{figure*}

\subsection{Extension to high-resolution digital temporal holography}

We demonstrate now that a simpler version of our heterodyne time microscope can be used to measure ultrafast dynamics of optical fields in single shot.  A spatial hologram is obtained by superposition of a reference light wave and of the light coming from an object. It contains all the phase and amplitude information characterizing the object under study. In digital holography, it is possible to recompose numerically the object by applying an appropriate paraxial diffraction operator to the measured hologram ~\cite{Schnars}. Transposing this idea from the spatial domain to the temporal domain, we have removed the Treacy compressor from the initial setup. Thus, our device is now conceptually equivalent to a digital temporal holography device and we call it SEAHORSE (Spatial Encoding Arrangement with Hologram Observation for Recording in Single shot the Electric field).

The 2D pattern recorded with the camera is the temporal anolog to a spatial hologram  contains phase and amplitude information of the initial field. As in digital (spatial) holography \cite{Schnars}, we can numerically retrieve the input signal by simply computing the propagation of the recorded electric field (here by applying an adequate dispersion $\beta$, see Sec.~Methods).

From the theoretical point of view, in the SEAHORSE the value of $\beta$ has to be exactly opposite to the dispersion experienced by the chirped pulse at 800 nm.

We have recorded the hologram corresponding to two 80~fs-pulses  separated by $\sim 2.7$~ps (see Sec. Methods). As the spectrum of the two pulses is extremely broad ($\sim 15$THz), the 2D hologram, [Fig.~\ref{fig:5}.(b)] and the corresponding  evolution of the intensity [Fig.~\ref{fig:5}.(c)] and of the phase [Fig.~\ref{fig:5}.(d)] are rather complex. However, using a GVD coefficient $\beta=-0.22$~ps$^2$ that exactly corresponds to the dispersion previously induced by the Treacy compressor, the numerical propagation of the field leads to two extremely short peaks that are separated by  $2.76$~ps. Remarkably, the pulse width is $\approx 80$~fs that correspond to the the distance between two pixels of the camera. This value can be considered as  representing the temporal resolution of our temporal digital holograpy device.

We want to point out that the effective resolution of the SEAHORSE ($<100$fs) is better than the resolution of the HTM. In the case of the time microscope, the resolution is limited by the aberrations due to third order dispersion of the Treacy compressor \cite{Backus:98}. The minimum temporal width that is  observable with our HTM depends on the signal spectral width (as expected from aberration studies \cite{Bennett:00}) and is around $200$~fs. SEAHORSE is somehow simpler than the usual time lens devices because the first dispersive element is not needed. Moreover SEAHORSE holographic reconstruction  appears as a promising way to improve the resolution of time-lenses, by strongly reducing the aberrations, as shown here, or -- potentially -- even manage the aberrations a posteriori.\\

\section*{Discussion}

In this Article, we have demonstrated two novel and complementary techniques that allow  the single-shot recording of amplitude and phase of irregular waves with a high temporal resolution ($\sim 80$fs) over a long time window ($\sim 40$~ps). The key point is a completely new combination  of an heterodyne technique \cite{Broaddus:10, Dorrer:06} and of the so-called time lens strategy that provide large temporal windows \cite{Kolner:89, Suret:16,Narhi:16}

We believe that our heterodyne time lens and time holography devices will represent complementary  tools to single shot versions of  fast-detection devices such as FROG~\cite{Trebino} or
SPIDER~\cite{Dorrer:99},  suitable for the measurement of complex and non reproductible optical signals. The comparison between  techniques  introduced in this paper and already-existing single shot phase and amplitude measurement devices ({\it e.g.} time stretch \cite{Herink:17,Mahjoubfar:17}, XFROG~\cite{Trebino,Wong:14}, SEA-SPIDER~\cite{Kosik:05,Wyatt:06}, SEA-TADPOLE~\cite{bowlan2006crossed} or STARFISH~\cite{alonso2010spatiotemporal})  represents a complex question that is out of the scope of the present paper and that deserves further  analysis.

We believe that the combination of large temporal window together with subpicosecond resolution represents one of the key-advantages of the setups presented in this Article. Moreover, our HTM and our time holography system do not require iterative or complex algorithm.

From the fundamental point of view, our HTM has unveiled the special evolution of the phase (and  amplitude) of rogue waves in integrable turbulence, and in particular their  compatibility with the expected  PS \cite{Kibler:10, Bertola:13, Dudley:14,Toenger:15}. Our results open the way to novel and fundamental investigations of nonlinear propagation of random waves, of the emergence of optical rogue waves  \cite{Solli:07, Akhmediev:09, Akhmediev:09b,Mussot:09, Suret:16,Narhi:16} and more generally of  nonlinear statistical optics~\cite{Picozzi:14}.

\newpage

\section*{Methods}


{\bf Heterodyne time microscope}\\

For the single-shot acquisition of phase and amplitude of sub-picosecond optical fields, heterodyne technique is implemented in an upconversion time-microscope composed  ~\cite{Bennett:99} of a {\it time lens} and of a single shot spectrometer very similar to the one described in Ref.~\cite{Suret:16}. The heterodyne time microscope encodes the temporal change of an interference pattern onto the spectrum of a chirped pulse. The fringe pattern is imaged onto the vertical axis of a sCMOS camera and the spectrum is imaged on the horizontal axis of the camera using  a 1800~lines/mm grating. The 2D pattern is recorded in single shot with $512 \times 100$ pixels at the  repetition rate of the $800$ nm pump laser ($1$~kHz).


We use an upconversion time-lens as in Ref.~\cite{Kolner:89}
  (see Fig.~\ref{fig:1}). A  1~mm-long  BBO crystal has cut for noncollinear
type-I SFG is pumped by a 800~nm chirped pulses   that are provided by an amplified Titanium-Sapphire laser (Spitfire from  Spectra Physics). The laser emits 2~mJ, 40~fs pulses with a spectral bandwidth of about 25~nm) at 1~kHz repetition rate, and only 20~$\mu$J are
  typically used here. For inducing the required normal dispersion on the  800~nm pulses we simply adjust the amplifier's output compressor
  dispersion. The dispersion was fixed to $0.21~$ps$^2$, leading to chirped pulses of duration of about $25~$ps.

Before being focused inside the BBO crystal, the 1560~nm signal under investigation experiences anomalous dispersion in a classic Treacy grating compressor (see Fig.~\ref{fig:1}). The 1560~nm grating compressor  is made with two 600~lines/mm  gratings, operated at an angle of incidence of 49~degrees, and whose planes are separated by
38.5~mm.  
Then, it is combined with a single-frequency reference beam at 1561~nm
by using a non polarizing cube. The single-frequency reference beam is
emitted by a tunable laser source having a narrow linewidth of  $\sim
300$ kHz (APEX AP3350A). The coherence time scale of  the reference
source ($\sim 3 \mu$s)  is much higher than the temporal windows of recording ($\sim 40$~ps) so that the phase of the reference can  be considered  as being constant over each observation window.  The reference light is amplified by using an Erbium-doped fibre amplifier (Keopsys) to a power of a few Watts. 

The 800~nm beam, the reference beam and the signal beam are designed
to assume a transverse elliptic shape by using cylindrical lenses. The
horizontal  waist diameter of the three beams is typically $50 \,\mu$m. The vertical waist of the 1560~nm beams are typically  $500 \,\mu$m and the vertical waist of the 800~nm beam is typically $1$~mm. An angle $\sim 10$~mrad is  adjusted in the vertical plane between the signal under investigation and the monochromatic reference in order to get several horizontal interference fringes in the observation plane.

In order to reject the 800~nm and 1560~nm radiations and to  illuminate the camera with only the SFG at $529$~nm, an iris and a 40-nm bandpass filter around 531~ nm (FF01 531/40-25 Semrock) is placed after the crystal. An achromatic lens with 200~mm focal length collimates the 529~nm light after the BBO crystal. The camera is a sCMOS Hamamatsu Orca flash 4.0 V3 (C13440-20CU), equiped with a 80~mm lens (Nikkor Micro 60~mm f/2.8 AF-D). The objective is focused at infinity and the waist of the SFG in the BBO crystal is imaged on the camera sensor. The camera is synchronized on the 800~nm laser pulses, and the integration time is
adjusted to 1~ms, thus enabling single-shot operation of the time-microscope.


As in other time-lens systems~\cite{Kolner:89,Bennett:99,  Foster:08}
high resolution requires proper adjustement of  the dispersion produced by the 1560~nm compressor. This is conceptually analog to the focus which is required in conventional microscope, and performed by adjusting the object-objective distance. We adjust the distance between the two gratings of the Treacy compressor by minimizing the width of the image of femtosecond pulses (see \cite{Suret:16} for the details of the procedure). The best temporal resolution is achieved for a dispersion value of $\beta \simeq -0.21$ps$^2$. This value is very close to the theoretical expectation, i.e., exactly opposite to dispersion experienced by the 800nm pulse. For all results presented in this paper, the temporal resolution of the HTM is $~200$~fs FWHM, and the field of view is around 40~ps.\\


{\bf Partially coherent light}\\

The partially coherent light used in the experiments reported in Fig.~\ref{fig:2} is generated by an Erbium fibre broadband Amplified Spontaneous Emission (ASE) source (Highwave). This broadband light is spectrally filtered (with programmable shape and linewidth)
using a programmable optical filter (Waveshaper 1000S, Finisar). The filtered light is then
amplified by using an Erbium-doped fibre amplifier (Keopsys).
In the experiments reported in Fig.~\ref{fig:4}, the 
partially-coherent light obtained in this way is launched inside a single-mode polarization maintaining fibre (Fibercore HB-1550T),  having either 100~m or 400~m length. The measured group velocity (second order) dispersion coefficient of the fibre is $\beta_2=-21$~ps$^2$km$^{-1}$. For a given spectral width, the power of the light beam launched inside the fibre is controlled using a
half-wavelength plate and a polarizing cube.\\


{\bf Time calibration of the HTM}\\

Time calibration of the HTM is performed by using series of two laser pulses that are separated by a well-calibrated time delay. The  pulses are emitted by a $1.5$ $\mu$m mode-locked laser (from Pritel) and have $\sim 1$~ps temporal width. Those laser pulses propagate inside a polarization maintaining fiber where they experience differential group delay depending on their state of polarization. The time separation ($5.6$ ps) between the two pulses is accurately measured from spectral interferometry. Observation
of the two pulses on the sCMOS camera then permits the accurate calibration of the time axis
and to measure that one camera pixel corresponds to a time duration of $78$ fs. \\


{\bf Data processing-Phase and power measurement}\\

Intensity along one vertical line at the position $x$ is given by the beating between the signal beam and the reference beam. It can be presented in the form: 

\begin{equation}\label{eq:I}
  \begin{split}
I&(x,y) = I_r(x,y) + I_s(x,y) \\
&+ \sqrt{I_r(x,y)\,.\, I_s(x,y)}\cos[k_y y + \phi (x) ]
\end{split}
\end{equation}

$I_r(x,y)$ represents the transverse intensity profile that is detected by the sCMOS camera  in absence of the signal beam and $I_s(x,y)$ represents the transverse intensity profile that is detected by the sCMOS camera  in absence of the reference beam. In the analysis presented in this paper, we neglect the fluctuations of the reference power. We thus compute $I_r(x,y)$ from additionnal experiments in which we record the SFG between the pump and the references beams (without the signal).

In the experiments, we have chosen the angle between the reference and
the signal beams  in order to oberve a sufficiently large number of fringes so that $ \int \sqrt{I_r(x,y)\,.\,I_s(x,y)}\cos[k_y y + \phi(x)] \, dy \simeq 0$. For each frame,  we thus simply compute the optical power $P(x)$ of the signal by using the formula :

\begin{equation}
    P(x)=\int I(x,y) \, dy - \int I_r(x,y)\, dy
\end{equation}

Note that here $P$ has arbitrary unit. In our study about partialy coherent waves (Figs.,~\ref{fig:2}, ~\ref{fig:4} and ~\ref{fig:5}) we compute the average $\langle P(x) \rangle $ from  $5 \time 10^4$ frames. We then display $P(x)/\langle P(x) \rangle $ or $[P(x)/\langle P(x) \rangle] \times P_0$ where $P_0$ is the averaged power launched inside the fiber. We remind that one pixel in $x$ corresponds to $78$~fs.\\

The relative phase between the signal and the reference $\phi(x)$ is simply given by the position of the maxima of the interference fringes [see Eq.(\ref{eq:I})]. $\phi(x)$ can be easily determined by the FFT of $I(x,y)$ in the variable $y$. $\phi(x)$ is given by the phase (complex argument) of the isolated spectral peak associated with the pattern that oscillates at
the frequency $k_y$ along the $y$ direction~\cite{Kreis:86}.\\ 


{\bf Time holography}

In our time holography setup, the signal under interest no longer experiences dispersion
by being injected in the Treacy compressor. In a conventional microscope, this amounts
to place the object under study very close to the microscope objective. With this
change, we record the complex field $\psi_H(t)$ associated with the temporal hologram
instead of the temporal image of the signal under investigation. 

The time evolution of the signal under investigation $\psi(t)$ can be obtained
by applying an appropriate dispersion operator in the Fourier space, i.e. 

\begin{equation}
\tilde{\psi} (\omega) = \tilde{\psi}_H (\omega) \; e^{i\beta \, \omega^2 \, /2}
  \label{eq:beta}
\end{equation}
where $\tilde{\psi}_H (\omega)$ and $\tilde{\psi} (\omega)$ are the Fourier transforms of $\psi_H(t)$ and of $\psi(t)$, respectively. From the theoretical point of view, the value of $\beta$ must be exactly equal to the group velocity dispersion (GVD) characterizing the chirped pump pulse used in the time lens.\\

The pair of ultra-short pulses used to demonstrate the capabilities of
the time holography is produced by a femtosecond Erbium Laser (ELMO,
from Menlo Systems  GmbH) equipped with a Michelson interferometer. The output
pulse duration at the 1553 nm central wavelength is $\leq 80$ fs and the
repetition rate is 100 MHz. The energy  of one pulse launched in the BBO crystal is typically $\sim 0.2$~nJ. In order to create the double pulse signal we used a Michelson
interferometer with tunable optical path difference. We fixed the
delay  between two pulses at  the value $\sim 2.7$ ps (measured with the full HTM including the Treacy compressor).


{\bf Numerical simulations of the 1D-NLSE}

Numerical simulations presented in Fig.~\ref{fig:3} and in Fig.~\ref{fig:5}
are performed by integrating the 1D-NLSE :
\begin{equation}
  \label{eq:NLS1D}
  i\frac{\partial \psi}{\partial z}=-\frac{\beta_2}{2}\frac{\partial^2 \psi}{\partial t^2}+\gamma|\psi|^2\psi.
\end{equation}
$\psi$ represents the complex envelope of the electric field, normalized so that $|\psi|^2$ is the
optical power. $z$ is the longitudinal coordinate measured along the fibre, and $t$ is the
retarded time. $\beta_2=-22$~ps$^2$km$^{-1}$
is the second-order dispersion coefficient of the fibre and
$\gamma$ is the Kerr coupling coefficient. From the comparison between
the optical spectra measured in the experiments and the ones computed from the
numerical simulations, we estimate that the Kerr coefficient is
$\gamma\sim 2.4$~W$^{-1}$km$^{-1}$.  All numerical integrations are performed
using splitstep pseudo-spectral methods.

In numerical simulations presented in this letter, we neglect linear
losses ($\simeq 0.5$dB in  the 400m-long PMF.) and stimulated Raman scattering. These
approximations provide precise and quantitative agreement between
experiments and numerical simulations at moderate powers ($<2$W).

\section*{Acknowledgments}

This work has been partially supported by the Agence Nationale de la
Recherche through the LABEX CEMPI project (ANR-11-LABX-0007) and the
OPTIROC project (ANR-12-BS04-0011 OPTIROC)  and by the Ministry of
Higher Education and Research, Nord-Pas de Calais Regional Council and
European Regional Development Fund (ERDF) through the Contrat de
Projets Etat-Région (CPER Photonics for Society P4S).  The authors are
grateful to Arnaud Mussot the photonics group of the PhLAM for
fruitful discussions and for providing the ps laser. The authors also gratefully aknowledge MENLO for proving the femtosecond laser used for the time-holography measurements. The authors thank Nunzia Savoia for the everyday work on the femto laser
and Rebecca El Koussaifi, Cl\'ement Evain and Marc Le Parquier for their crucial contribution in the development of the time lens. The authors thank Pascal Szriftgiser from PhLAM for giving us access to some specific equipments.

\section*{Author  contributions}


All authors contributed to the design and the realization of the heterodyne time microscope and time holography devices.  All the authors participated to the data acquisition that has been essentially performed by A.T. All authors participated to data analysis, numerical simulations and have written the manuscript.

\end{document}